\documentclass[12pt]{article}
\usepackage{amsmath}
\usepackage{amsfonts}
\usepackage{amssymb}
\usepackage[T2A]{fontenc}
\usepackage[cp1251]{inputenc}
\usepackage{graphicx}

\begin{document}
\centerline{\Large\sc Dynamical Symmetry Breaking on a Cylinder}
\vspace{3mm}
\centerline{\Large\sc in Magnetic Field}
\vspace{8mm}

\hspace{3cm}

{\small \bf

\quad \quad \quad \quad \quad \quad A. V. Gamayun $^{(a,c)}$}
\footnote{E-mail address: gamayun@bitp.kiev.ua}
\quad
{\small \bf
and \quad E.V. Gorbar $^{(b,c)}$}
\footnote{E-mail address: egorbar@uwo.ca}

\vspace{8mm}

{\small\sl

\quad \quad \quad \quad (a) Institute of Physics and Technology of
National Technical University,
\\ \centerline{"Kiev Polytechnic
Institute" Peremogy Av., 37, 03056 Kiev, Ukraine}}

\vskip 4mm

{\small\sl

\quad \quad \quad \quad (b) Department of Applied Mathematics, University of Western Ontario,\\
\centerline{London, Ontario N6A 5B7, Canada}}

\vskip 4mm

{\small\sl

\quad \quad \quad \quad (c) Bogolyubov Institute for Theoretical Physics, 03143 Kiev, Ukraine}

\vspace{4cm}

\begin{abstract}
\vspace{3mm} We study dynamical symmetry breaking on a cylinder in
external magnetic field parallel to the axis of cylinder when
magnetic field affects the dynamics of fermions only through the
Aharonov-Bohm phase. We find that unlike  other previously studied
cases magnetic field in our case counteracts the generation of
dynamical fermion mass which decreases with magnetic field. There
exists also a purely kinematical contribution to the fermion gap
which grows linearly with magnetic field. Remarkably, we find that
the total fermion gap, which includes both the dynamical and
kinematical contributions, always increases with magnetic field
irrespectively the values of coupling constant and the radius of
cylinder. Thus, although the dynamical mass is suppressed,
external magnetic field does enhance the total fermion gap in the
spectrum.
\end{abstract}

\newpage

\section{Introduction}

\vspace{3mm}

It is well known that external magnetic field is a strong catalyst
of dynamical symmetry breaking leading to the generation of
dynamical fermion mass even at the weakest attractive interaction
between fermions \cite{GMSh} (for earlier consideration of
dynamical symmetry breaking in a magnetic field, see
\cite{magnetic1, magnetic2}). The applications of this effect were
considered in condensed matter and cosmology (for reviews see
\cite{reviews}). Recently, the idea of magnetic catalysis was used
for explanation of the experimentally observed magnetic field
driven metal-dielectric phase transition in pyrolytic graphite
\cite{graphite}.

As well known, carbon nanotubes (see, e.g., \cite{NT}) are
essentially graphite sheets wrapped on a cylinder. Experimentally
produced carbon nanotubes have small radii and their long wave
length excitations are adequately described by a (1+1)-dimensional
effective quantum field theory whose spatial content is a
$\mathbf{R}^1$ space because only the lowest mode for fermions on
the circle is retained. Nonetheless, the cylinder geometry of
carbon nanotubes presents an interesting setup from the viewpoint
of dynamical symmetry breaking in external constant magnetic field
in a (2 + 1)-dimensional spacetime with nontrivial topology. We
study this problem in the present paper. It is clear that two
directions of magnetic field are distinguished. The first one is
when magnetic field is perpendicular to the axis of cylinder and
the second when it is parallel. Obviously, since the normal
projection of magnetic field varies with angle, the first case
presents an inhomogeneous problem whose solution is difficult to
find. The case where magnetic field is parallel to the axis of
cylinder is much more tractable and was already studied in some
detail in \cite{KSH}. Classically, magnetic field parallel to the
axis of cylinder does not affect at all the motion of charged
particles on the surface of cylinder. However, this is not true in
quantum mechanics where the presence of magnetic field leads to
the appearance of the Aharonov--Bohm phase \cite{AB}. We would
like to add also that, in general, one can relax the requirement
that external magnetic field is constant. It is enough to require
only that the normal component of magnetic field be equal to zero
on the surface of cylinder. Magnetic field through the transverse
section of cylinder can be arbitrary and what matters is only the
total magnetic field flux through the transverse section of
cylinder.

To study dynamical symmetry breaking on a cylinder in parallel
magnetic field, we consider the following Gross-Neveu type model
\cite{GN} with $N$ flavors:
\begin{equation}
\mathcal{L}=\sum^{N}_{k=1}\bar{\psi_{k}}i\gamma^{\mu}D_{\mu}\psi_{k}+
\frac{G}{2N} (\bar{\psi}\psi)^2, \label{w0}
\end{equation}
where $D_{\mu} = \partial_{\mu} -ieA_{\mu}^{ext}$ and
$\bar{\psi}\psi = \sum_{i=1}^{N} \bar{\psi}_i\psi_i$, $N$ the
number of flavors. According to \cite{Appelquist}, we consider the
four-component spinors corresponding to a four-dimensional
(reducible) representation of Dirac`s matrices
\begin{equation}
\gamma^{0}=\left(%
\begin{array}{cc}
  \sigma^3 & 0 \\
  0 &  -\sigma^3 \\
\end{array}%
\right)\,\,\,,\,\,\,\,\,\,
\gamma^{1}=\left(%
\begin{array}{cc}
  i\sigma^1 & 0 \\
  0 &  -i\sigma^1 \\
\end{array}%
\right)\,\,\,,\,\,\,\,\,\,\,
\gamma^{2}=\left(%
\begin{array}{cc}
  i\sigma^2 & 0 \\
  0 &  -i\sigma^2 \\
\end{array}%
\right).
\end{equation}
The Lagrangian (\ref{w0}) is invariant under the discrete
transformation $\psi \rightarrow e^{i\frac{\pi}{2}\gamma^5}\psi$, where
$$
\gamma^5 = \left(%
\begin{array}{cc}
  0  &  I \\
  I &  0 \\
\end{array}%
\right).
$$
Obviously, the mass term breaks this symmetry since
$$
\bar{\psi} \psi \rightarrow - \bar{\psi}\psi.
$$
We consider the model with breaking of discrete symmetry in order
to avoid strong quantum fluctuations due to massless
Nambu--Goldstone fields which, if present, would destroy the
mean-field solution that we find.

\section{Dynamical mass generation}

\vspace{3mm}

Introducing an auxiliary $\sigma$ field, the Lagrangian (\ref{w0})
can be equivalently represented in the following way:

\begin{equation}
\mathcal{L}=\sum^{N}_{k=1}(\bar{\psi_{k}}i\gamma^{\mu}D_{\mu}\psi_{k}-
\sigma\bar{\psi_{k}}\psi_{k})-\frac{N}{2G}\sigma^2\,.
\end{equation}

Integrating over the fermion fields, we obtain the
following effective action for the $\sigma$ field in the
$\frac{1}{N}$ approximation:
\begin{equation}
\Gamma(\sigma)= -\frac{N}{2G}\int \sigma^2d^3x - iNTr Ln
(i\gamma^{\mu}D_{\mu}-\sigma).
\end{equation}
If we choose the $x$ axis along the axis of cylinder and the $y$
axis along the circumference, then, mathematically, cylinder is
a $\mathbf{R}^1\times\mathbf{S}^1$ space or
compactified plane in the $y$ direction, i.e. $y \sim y+2\pi R$,
where $R$ is the radius of cylinder. Obviously,
$\mathbf{R}^1\times\mathbf{S}^1$ is a multiply-connected space
and, according to Hosotani \cite{Hosotani}, the nonzero component
$A_y$ of gauge field cannot be gauged away unlike the case of a
simply connected space. Equivalently, if we consider our space
$\mathbf{R}^1\times\mathbf{S}^1$ as a real cylinder in embedding
3D space, then the constant $A_y$ component is related to the flux of
constant magnetic field $\Phi=2\pi R A_y = \pi R^2 B$ through
the transverse section of cylinder \cite{LN}. Further,
\begin{equation}
Tr\ln(i\gamma^{\mu}D_{\mu}-\sigma)= -Tr\int^{\sigma}_{0} ds
\frac{1}{i\gamma^{\mu}D_{\mu}-s}= 4\int d^3x \int^{\sigma}_{0} sds
G(\mathbf{x},\mathbf{x};s)
\end{equation}
and, consequently, the effective potential is equal to
\begin{equation}
V(\sigma) = \frac{N\sigma^2}{2G} + 4Ni\int^{\sigma}_{0} sds
G(\mathbf{x},\mathbf{x};s),
\end{equation}
where $G(\mathbf{x},\mathbf{x^{\prime}};s)$ is Green`s function
which satisfies the following equation:
\begin{equation}
(\frac{\partial^2}{\partial t^2}-\frac{\partial^2}{\partial
x^2}-(\frac{\partial}{\partial
y}-ieA)^2+s^2)G(\mathbf{x},\mathbf{x^{\prime}};s)=\delta(x-x^{\prime}),
\label{w1}
\end{equation}
where $A_y=\frac{RB}{2}$. Obviously, in Euclidean space,
\begin{equation}
G(\mathbf{x},\mathbf{x};s)= -i\int \frac{d^2p}{(2\pi)^2L} \sum_n
\frac{1}{p^2+(\frac{2\pi }{L})^2(n-\phi_{\|})^2+s^2}\,, \label{w3}
\end{equation}
where $L=2\pi R$, $\phi_{\|}=\Phi/\Phi_0 = \frac{eA_yL}{2\pi}$, and
$\Phi_0=2\pi/e$ is the elementary magnetic field flux. Since the sum in (\ref{w3})
is over all integer $n$, it suffices to consider $\phi_{\|}$ on
the interval $[0,\frac{1}{2}]$. To evaluate the sum in (\ref{w3}), we
transform it into an integral in complex plane $\omega$
\begin{equation}
\sum_n \frac{1}{(p^2+s^2)(L/2\pi)^2+
(n-\phi_{\|})^2}=\frac{1}{2\pi i}\int_{C}
\frac{d\omega}{1-e^{2\pi(\omega+i\phi_{\|})}}\frac{-2\pi}{(p^2+s^2)(L/2\pi)^2-\omega^2}\,,
\end{equation}
where the contour $C$ runs around poles of the function
$(1-e^{2\pi(\omega+i\phi_{\|})})^{-1}$. Then we can deform contour
and present the last integral as sum over two residues at
$\omega=\pm L/2\pi (p^2+s^2)^{1/2}$. As result, we have
\begin{equation}
G(\mathbf{x}, \mathbf{x}; s)=-i\frac{1}{4\pi}\int
\frac{pdp}{(p^2+s^2)^{1/2}}\frac{\sinh(L(p^2+s^2)^{1/2})}{\cosh(L(p^2+s^2)^{1/2})-\cos(2\pi\phi_{\|})}=
\end{equation}
$$ -\frac {i}{4\pi L}\ln
\frac{\cosh(L(\Lambda^2+s^2)^{1/2})-\cos(2\pi\phi_{\|})}{\cosh(Ls)-\cos(2\pi\phi_{\|})}\,.
$$
Finally, the effective potential is
\begin{equation}
V(\sigma)=\frac{N\sigma^2}{2G}-\frac{N}{\pi L}\int^{\sigma}_{0}
\ln
\frac{\cosh(L(\Lambda^2+s^2)^{1/2})-\cos(2\pi\phi_{\|})}{\cosh(Ls)-\cos(2\pi\phi_{\|})}
sds,  \label{potential}
\end{equation}
where $\Lambda$ is cut-off. The gap equation which follows from $\frac{dV(\sigma)}{d\sigma}=0$ is
\begin{equation}
\frac{\sigma}{G} = \frac{\sigma}{\pi L}\ln
\frac{\cosh(L(\Lambda^2+\sigma^2)^{1/2})-\cos(2\pi\phi_{\|})}{\cosh(L\sigma)-\cos(2\pi\phi_{\|})}\,.
\label{w4}
\end{equation}
Consequently, the nontrivial solution for $\Lambda \to \infty$ is
\begin{equation}
\sigma=\frac{1}{L} \cosh^{-1} (\frac{e^{L(\Lambda-\frac{\pi}{G})}}{2}+\cos(2\pi\phi_{\|}))\,,
\label{solution}
\end{equation}
which obviously decreases with $\phi_{\|}$ (see Fig.1). We would
like to note that our effective potential and the solution of gap
equation agree with the corresponding results obtained in
\cite{KSH}. By using (\ref{potential}), it is not difficult to
show that the nontrivial solution always has lower energy than the
trivial solution $\sigma=0$.
\begin{center}
\includegraphics*[scale=0.8]{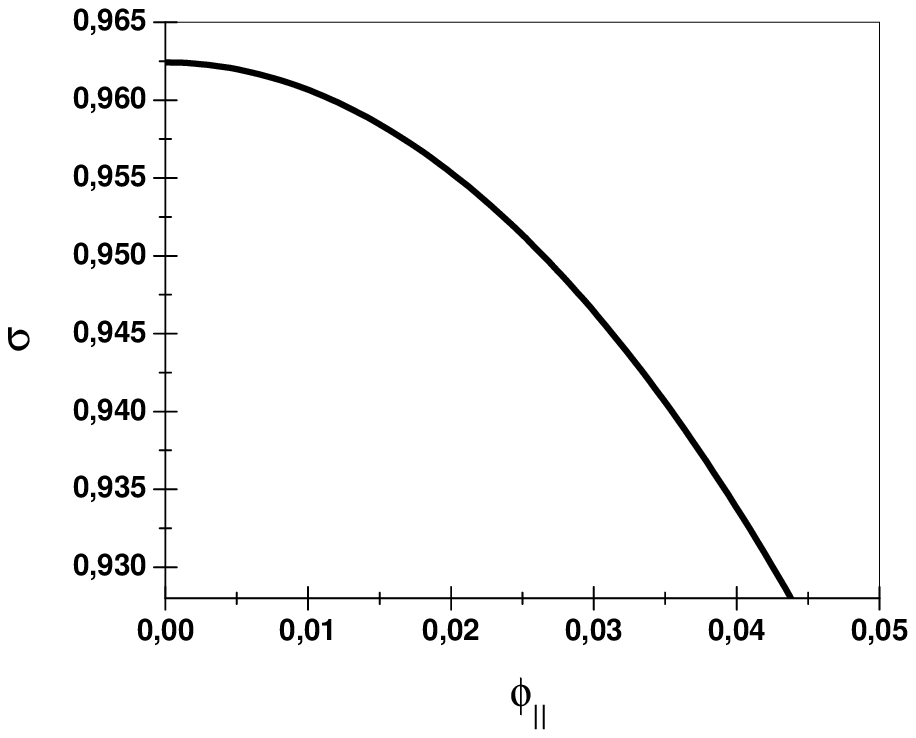}
\vspace{3mm}

FIG.1. The dependence of $\sigma$ on $\phi_{\|}$.
\end{center}
\vspace{3mm}
This result corresponds to an unusual situation where
external constant magnetic field inhibits rather than assists
dynamical symmetry breaking. According to \cite{GMSh}, constant
magnetic field in infinite flat space leads to dimensional
reduction by two units $D + 1 \to (D - 2) + 1$ in the infrared
region for fermions that strongly assists dynamical symmetry
breaking. Magnetic field in our case does not influence the
spatial motion of fermions at all, therefore, the dimensional
reduction is absent. The only dynamical effect that magnetic field
has in our case is through the appearence of the Aharonov--Bohm
phase. Since this phase increases energy of fermions, we obtain
that magnetic field in our case counteracts dynamical symmetry
breaking unlike all other known cases. Remarkably, this result is
consistent with a more general idea that external magnetic field
enhances the fermion gap in the spectrum. The point is that there
is an additional purely kinematic contribution
$\frac{2\pi\phi_{\|}}{L}$ (see (\ref{w3})) due to the
Aharonov--Bohm phase to the total fermion gap and as we show below
the total gap
$\sigma_{tot}^2=\sigma^2+(\frac{2\pi\phi_{\|}}{L})^2$ increases
with $\phi_{\|}$. To prove this, let us consider the function
$$
f(x)=(\cosh^{-1}(a+\cos x))^2+x^2,
$$
where $x=2\pi\phi_{\|}$ takes values on the interval $[0,\,\pi]$ and $\xi = a + \cos x > 1$, otherwise, the gap
equation (\ref{w4}) does not have a nontrivial solution.
The derivative of this function with respect to $x$ is equal to
$$
\frac{df(x)}{dx}=2x(1-\frac{\sin x}{x}\frac{\ln(\xi+\sqrt{\xi^2-1})}{\sqrt{\xi^2-1}})\,.
$$
Further, taking into account that
$$\frac{\sin x}{x}\leq 1\,\,\,\,\,\,\,\,\, \mbox{and}\,\,\,\,\,\,\,\,\,
\frac{\ln(\xi+\sqrt{\xi^2-1})}{\sqrt{\xi^2-1}} < 1\,,
$$
we find that this derivative is positive
$$
\frac{df(x)}{dx} > 0\,.
$$
Thus, although the dynamically generated fermion mass decreases with $\phi_{\|}$, the total fermion gap
increases with $\phi_{\|}$.
Graphically, the dependence of $\sigma_{tot}$ on $\phi_{\|}$ is depicted in Fig.2.
\begin{center}
\includegraphics*[scale=0.8]{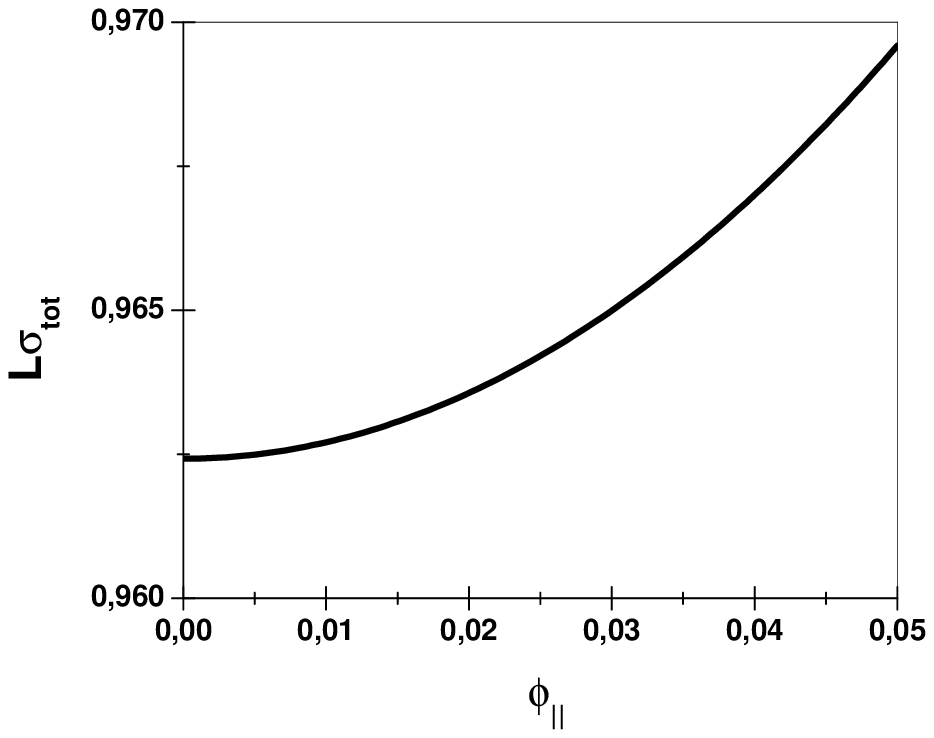}
\vspace{0mm}

FIG.2. The dependence of $\sigma_{tot}$ on $\phi_{\|}$.
\end{center}
\vspace{3mm}

Finally, let us find the critical coupling constant that separates the symmetrical phase from the phase with broken symmetry.
As follows from (\ref{solution}), the critical coupling constant is
\begin{equation}
G_c(\phi_{\|}) = \frac{\pi L}{\Lambda L - \ln[2(1-\cos(2\pi \phi_{\|}))]}\,.
\label{critical}
\end{equation}
Once again we see that external magnetic field counteracts
dynamical symmetry breaking since as follows from (\ref{critical})
the critical coupling constant increases with $\phi_{\|}$ that
means that we need stronger attraction in order to break symmetry.
For $\phi_{\|} \to 0$, the critical coupling constant goes to zero
$G_c \to 0$, i.e. symmetry is always broken in this case.By using
(\ref{solution}) one can show that the phase transition with
respect to the magnetic field flux is the mean field second order
phase transition.

\section{Conclusions}

\vspace{3mm}

In this paper we have investigated dynamical symmetry breaking on a cylinder in external magnetic field parallel to
the axis of cylinder. This problem may be relevant for certain condensed matter systems. In fact, we were
inspired by the geometry of carbon nanotubes. On the other hand, our problem can be considered as the problem of dynamical
symmetry breaking in a multiply-connected space, where gauge field has a nonzero
constant vacuum expectation value \cite{Hosotani}, which cannot be gauged away unlike the case of a
simply connected space.

By studying a relatively simple Gross--Neveu type model on a
cylinder in external magnetic field, we find that unlike all other
known cases external magnetic field counteracts the generation of
dynamical fermion mass, i.e. larger magnetic fields correspond to
smaller dynamical fermion masses. There exists also an additional
purely kinematic contribution to the fermion gap in this problem,
which increases with magnetic field. Remarkably, we find that the
total fermion gap, which includes both the dynamical and
kinematical contributions, always increases with magnetic field
irrespectively the values of coupling constant and the radius of
cylinder. Thus, although our analysis shows that external magnetic
field in spaces with nontrivial topology does not always assist
dynamical symmetry breaking unlike the case of flat spaces with
trivial topology \cite{GMSh}, our results are consistent with a
more general idea that external magnetic field increases the
fermion gap in the spectrum.

We would like to note that one can gauge away the vector potential
in our problem but in this case the Aharonov--Bohm phase will
reveal itself in the boundary conditions for fermions on the
circle. The cases of periodic and antiperiodic boundary conditions
for fermion fields were considered in \cite{boundary}, where
dynamical chiral symmetry breaking was studied in a spacetime
$\mathbf{R^3} \times \mathbf{S^1}$ in external magnetic
field.\footnote{The role of periodic and antiperiodic boundary
conditions of fermion fields in the dynamical symmetry breaking in
flat spaces with more general topology but without external
magnetic field was considered in \cite{topologies}.} It is clear
that the Aharonov--Bohm phase defines arbitrary boundary
conditions for fermions on the circle and allows to interpolate
smoothly between the periodic and antiperiodic boundary
conditions. It is easy to check that our results on dynamical
symmetry breaking are consistent with the results of
\cite{boundary} (as well as \cite{topologies}), where it was found
that although for periodic boundary conditions (when the
Aharonov--Bohm phase is zero in our setup) a dynamical mass is
generated at the weakest attractive interaction between fermions,
the effect of antiperiodic boundary conditions (when the
Aharonov--Bohm phase attains its maximal possible value) is to
counteract the dynamical chiral symmetry breaking. Since magnetic
field considered in \cite{boundary} is perpendicular to the
spatial plane, unlike our case it cannot influence boundary
conditions for fermions on the circle. The paper \cite{FI} studied
dynamical chiral symmetry breaking in QED in a spacetime with the
topology $ \mathbf{R}^3 \times \mathbf{S}^1$ without external
magnetic field. In this case the Aharonov-Bohm phase for fermions
appears due to nonzero constant vacuum expectation value of the
electromagnetic vector potential that cannot be gauged away in a
multiply-connected spacetime. The Aharonov-Bohm phase in this
problem is not an external parameter like in our problem but a
dynamical variable that should be determined from the requirement
of the minimum of energy. Interestingly, it was found in \cite{FI}
that the state with the lowest energy corresponds to fermions with
antiperiodic boundary conditions. This disfavors dynamical fermion
mass generation, however, according to the results of our paper,
the total fermion gap which includes also the purely kinematic
contribution due to the Aharonov-Bohm phase is still larger than
the gap in the case of zero Aharonov-Bohm phase.

Finally, we would like to note that a (2 + 1)-dimensional free fermion model was investigated in \cite{induced}, where a nonzero chiral symmetry
breaking condensate was found due to the Aharonov--Bohm phase related to the singular magnetic vortex. In our opinion, it would be very interesting to
study dynamical symmetry breaking in this model by adding some interaction. Unfortunately, unlike the problem that we investigated in this paper the
problem considered in \cite{induced} is inhomogeneous and such a study would be very difficult to perform.

\vspace{5mm}

\centerline{\bf Acknowledgements}

\vspace{5mm}

We are grateful to V.P. Gusynin and V.A. Miransky for useful
discussions and critical remarks. E.V.G. thanks K.G. Klimenko for
bringing his attention to papers \cite{KSH}.The work of E.V.G. was
supported by the Natural Sciences and Engineering Research Council
of Canada.

\end{document}